\documentclass{ws-ijmpa}

\usepackage[super]{cite}
\usepackage{xcolor}
\usepackage[verbose,hypertexnames=false]{hyperref}
\usepackage{subcaption}
\usepackage{multirow}
\hypersetup{colorlinks=false,allbordercolors=blue,pdfborderstyle={/S/U/W 1}}

\begin{document}

\markboth{Boycha}{EGUP effects on KNBHQ}

%
\catchline{}{}{}{}{}
%

\title{EGUP effects on the thermodynamic properties of the Kerr-Newman black hole surrounded by quintessence
}

\author{Aheibam Boycha Meitei$^{1,\dag}$, Yenshembam Priyobarta
Singh$^{2,\dotplus} $,\\ T. Ibungochouba Singh$^{2,\ddagger}$, Irom Ablu Meitei$ ^{1,\ast}$ and Kangujam Yugindro Singh$ ^{1,\parallel} $}

\address{$ ^{1} $Department of Physics,\\ Manipur University, Canchipur 795003,\\ Imphal, Manipur, India.\\$ ^{2} $Department of Mathematics,\\ Manipur University, Canchipur 795003,\\ Imphal, Manipur, India.\\
$ \dag $aheibamboycha143@gmail.com\\$ \dotplus $priyoyensh@gmail.com\\$ ^{\ddagger} $ibungochouba@rediffmail.com (corresponding author)\\$ ^{\ast} $ablu.irom@gmail.com\\$ ^{\parallel} $yugindro361@gmail.com}

\maketitle

\begin{history}
\received{Day Month Year}
\revised{Day Month Year}
\accepted{Day Month Year}
\published{Day Month Year}
\end{history}

\begin{abstract}
In this paper, we investigate the effects of the Extended Generalized Uncertainty Principle (EGUP) on the thermodynamic quantities of the Kerr-Newman black hole surrounded by quintessence (KNBHQ). Additionally, we conduct a comparative analysis of the outcomes derived from the Generalized Uncertainty Principle (GUP) and the Extended Uncertainty Principle (EUP). The GUP is crucial in the context of the early universe, whereas the EUP is significant in the framework of the later universe. This analysis illustrates the variations in thermodynamic quantities, including Hawking temperature, heat capacity, Gibbs free energy, entropy and pressure of the Kerr-Newman black hole, as they evolve from the early universe to the latter universe under the influence of the dark energy model known as quintessence. The remnant mass, temperature and the stability of the KNBHQ are also studied.
\end{abstract}

\keywords{Extended Generalized Uncertainty Principle; Kerr-Newman black hole; Quintessence.}

\ccode{PACS numbers: 03.65.$-$w, 04.62.+v}

\section{Introduction}	
The non-decreasing property of the event horizon's area placed a critical restriction on the possible behavior of the black holes. The black hole's area closely resembles the behavior of a physical quantity called entropy\cite{1,2,3}. Following the discovery, Bekenstein\cite{4,5,6} emphasized that the event horizon's area was a measure of the entropy of the black hole. If a black hole possesses entropy, it must also have a temperature. But a body with a particular temperature must emit radiation at a certain rate. This radiation is now known as Hawking radiation and has become an important aspect in the study of the black hole. In the context of quantum mechanical effects, Hawking\cite{7,8} demonstrated that the radiation emitted is inherently thermal, which can be characterized by thermodynamic parameters such as temperature, entropy, and heat capacity. This has established a foundational basis for the study of black hole thermodynamics. Black hole thermodynamics illustrates the profound relationship between general relativity and quantum theory, highlighting the quantum nature of the black holes.

Beyond the Planck length $( l_{p})$, the smoothness of observed spacetime is not assured. This chaotic structure of spacetime at the microscopic level is referred to as quantum foam\cite{9}. To accommodate the  Planck length, the Heisenberg Uncertainty Principle (HUP) must be modified to the Generalized Uncertainty Principle (GUP)\cite{10,11,12}. Some authors discussed\cite{13,14,15,15a,15b,15c,15d} the thermodynamic behavior of black holes under the influence of GUP. The GUP is an effective way to realize the minimum observable length, which was predicted by various theories of quantum gravity and string theory\cite{16,17,18,19,20}. The most basic form of the generalized uncertainty relation can be expressed as\cite{21}
\begin{eqnarray}\label{eq1}
\Delta x\Delta p\geq\frac{\hbar}{2}\left[ 1+\beta(\Delta p)^{2}\right],
\end{eqnarray}
where $\beta=\beta_{0}\frac{l_{p}^{2}}{\hbar^{2}}$, $ \beta_{0} $ is a dimensionless parameter of order unity and $l_{p}=\sqrt{\frac{G\hbar}{c^{3}}}$ is the Planck length.

 As nature exhibits a preference for symmetry and duality, it is reasonable to consider that if a minimum fundamental length exists, there must also be a large fundamental length scale, $ L_{\ast} $, in our Universe. Therefore, the GUP is naturally extended to incorporate the significant fundamental length via a quadratic correction in the position of uncertainty\cite{22}
\begin{eqnarray}\label{eq2}
\Delta x\Delta p\geq\frac{\hbar}{2}\left[ 1+\frac{\eta(\Delta x)^{2}}{L_{\ast}^{2}}\right],
\end{eqnarray} 
  which we called an EUP, with $ \eta $ being another dimensionless constant parameter. Then, the effects of the EUP begin to be studied in the thermodynamics of the Friedmann–Robertson–Walker Universe\cite{23}, the geometry of de Sitter and anti-de Sitter spacetimes\cite{24}, shadow and lensing in the macro and microscopic realms\cite{25}, Rindler and cosmological horizons\cite{26}, and relativistic Coulomb potential\cite{27}. The EUP could contribute to dark matter effects\cite{28}. Numerous studies concerning black hole physics are also examined within the context of EUP\cite{29,30,31,32,33,34,35,36,37}.

The investigation of the most general form of the uncertainty principle through the integration of EUP and GUP is presented in Ref.\cite{38} and we called it the EGUP. Bolen and Cavaglia\cite{39} studied the thermodynamics behavior of Schwarzschild anti de Sitter black holes under the influence of EGUP.

The majority of energy in the Universe is comprised of some forms of dark energy, which exhibits gravitational self-repulsive properties, resulting in the accelerated expansion of the Universe\cite{40,41,42}. The primary candidates for this dark energy include the cosmological constant and quintessence. The cosmological constant is a static entity defined by an equation of state which is characterized by $ \omega=-1 $. In contrast, quintessence is a dynamic phenomenon with an equation of state represented as $ \omega=\frac{p}{\rho} $, where $ p $ denotes pressure and $ \rho $ signifies energy density, within the range of $ -1<\omega<-\frac{1}{3} $. In this paper, we investigate the effects of EGUP on the thermodynamic quantities of the KNBHQ.

The paper is organized as follows: In section 2, we provide a brief overview of EGUP. In section 3, we discuss the KNBHQ. In section 4, we investigate the effects of EGUP on the thermodynamic quantities of the KNBHQ, and a comparative analysis is done with the results derived from the GUP and EUP. Discussion and conclusions are given in section 5.

In this paper, we use natural unit system ($ \hbar=c=G=1 $).

\section{A brief overview of  EGUP}
In the early 1970s, Stephen Hawking, using the ideas of quantum field theory in curved spacetime, showed that black holes emit particles and the energy spectrum of the emitted particles is thermal\cite{7,8}. This indicates the quantum nature of black holes. Consequently, the emitted particle must comply with the fundamental commutation relation 
\begin{eqnarray}\label{eq3}
[\hat{x}_{\mu},\hat{p}_{\nu}]=i\delta_{\mu\nu},\  \ \mu,\nu=1,2,3.
\end{eqnarray}
The Heisenberg Uncertainty Principle (HUP), as derived from Eq. (3), is presented as follows
\begin{eqnarray}\label{eq4}
\Delta x \Delta p \geqslant \frac{1}{2} \delta_{\mu\nu}.
\end{eqnarray}
The existence of a minimal length of the order of Planck length, as predicted by various theories of quantum gravity, leads to a modified Heisenberg uncertainty principle (HUP) to the Generalized Uncertainty Principle (GUP) through a modified commutation relation 
\begin{eqnarray}\label{eq5}
[\hat{x}_{\mu},\hat{p}_{\nu}]=i\delta_{\mu\nu}[1+\beta \hat{p}^{2}],
\end{eqnarray}
where $ x_{\mu} $ and $p_{\nu}  $ are the position and momentum operators respectively. The expression of GUP is presented in Eq. (\ref{eq1}).

It is observed that Eq. (\ref{eq1}) lacks aesthetic appeal and does not exhibit a sense of equality, as the variables representing position $x$ and momentum $ p $ appear to have differing roles. Specifically, the square of $ \Delta p $ is present on the right-hand side of Eq. (\ref{eq1}), whereas the square of $ \Delta x $ is not included. This observation leads one to consider the potential for extending Eq. (\ref{eq1}) through the most general form of commutation relation
\begin{eqnarray}\label{eq6}
[\hat{x}_{\mu},\hat{p}_{\nu}]=i\delta_{\mu\nu}[1+\beta \hat{p}^{2}+\frac{\eta}{L_{\ast}^{2}}\hat{x}^{2}].
\end{eqnarray}
In this context, we examine the scenario where both $ \beta $ and $ \eta $ are greater than zero. If $ \eta=0 $, the Eq. (\ref{eq6}) simplifies to Eq. (\ref{eq5}) and subsequently reduces to GUP given in Eq. (\ref{eq1}). If $ \beta=0 $, we get the commutation relation
\begin{eqnarray}\label{7}
[\hat{x}_{\mu},\hat{p}_{\nu}]=i\delta_{\mu\nu}[1+\frac{\eta}{L_{\ast}^{2}}\hat{x}^{2}]
\end{eqnarray}
and degenerated into EUP given in Eq. (\ref{eq2}). Again, if we set $\eta=0 $ and $ \beta=0 $, we observe that Eq. (\ref{eq6}) simplifies to HUP.

Similarly, the most general form of the uncertainty principle, known as EGUP, can be deduced from Eq. (\ref{eq6}) as follows
\begin{eqnarray}\label{eq8}
\Delta x\Delta p\geq\frac{1}{2}\left[ 1+\beta(\Delta p)^{2}+\frac{\eta(\Delta x)^{2}}{L_{\ast}^{2}}\right].
\end{eqnarray} 
By solving Eq. (\ref{eq8}), we determine that the uncertainty in momentum $ \Delta p $ falls within the following range
\begin{eqnarray}\label{eq9}
\frac{\Delta x }{\beta}\Big[1-\sqrt{1-\beta\left( \frac{1}{(\Delta x)^{2}}+\frac{\eta}{L_{\ast}^{2}}\right) }\Big]\leqslant \Delta p \cr \leqslant \frac{\Delta x }{\beta}\Big[1+\sqrt{1-\beta\left( \frac{1}{(\Delta x)^{2}}+\frac{\eta}{L_{\ast}^{2}}\right) }\Big].
\end{eqnarray}

Similarly, the uncertainty in position $ \Delta x $ is also in the range
\begin{eqnarray}\label{eq10}
\frac{L_{\ast}^{2}\Delta p }{\eta}\Big[1-\sqrt{1-\frac{\eta}{L_{\ast}^{2}}\left( \frac{1}{(\Delta p)^{2}}+\beta\right) }\Big]\leqslant \Delta x \cr \leqslant\frac{L_{\ast}^{2}\Delta p }{\eta}\Big[1+\sqrt{1-\frac{\eta}{L_{\ast}^{2}}\left( \frac{1}{(\Delta p)^{2}}+\beta\right) }\Big].
\end{eqnarray}
The minimum uncertainty in momentum $ (\Delta p)_{min} $ and the minimum uncertainty in position $ (\Delta x)_{min} $ can be derived from Eqs. (\ref{eq9}) and (\ref{eq10}) respectively.

\section{Kerr-Newman black hole in the presence of quintessence}
In the Boyer-Lindquist coordinates $(t,r,\theta,\phi)$, the Kerr-Newman metric in the Kiselev quintessence is\cite{43}
\begin{eqnarray}\label{eq11}
ds^{2}=&-&\left( 1-\frac{2Mr-Q^{2}+\alpha r^{1-3\omega}}{\Sigma^{2}}\right) dt^{2}+\frac{\Sigma^{2}}{\Delta_{r}}dr^{2}\cr &-&\frac{2a \sin^{2}\theta (2Mr-Q^{2}+\alpha r^{1-3\omega})}{\Sigma^{2}}d\phi dt+\Sigma^{2}d\theta^{2}\cr &+& \sin^{2}\theta \left( r^{2}+a^{2}+a^{2}\sin^{2}\theta \frac{2Mr-Q^{2}+\alpha r^{1-3\omega}}{\Sigma^{2}}\right) d\phi^{2},
\end{eqnarray}
where 
\begin{eqnarray}\label{eq12}
\Delta_{r}&=&r^{2}-2Mr+a^{2}+Q^{2}-\alpha r^{1-3\omega},\cr
\Sigma^{2}&=&r^{2}+a^{2}\cos^{2}\theta.
\end{eqnarray}
$ M $, $ a $, $\alpha$ and $ Q $ represent the black hole mass, angular momentum per unit mass, quintessence parameter  and electric charge of the black hole respectively. The quintessence field parameter $ \omega $ defined by the equation of state $ \omega=\frac{p}{\rho} $, where $ p $ and $ \rho $ are the pressure and density is in the range $-1<\omega<-\frac{1}{3}  $.

The rotating characteristics of the KNBHQ gives rise to a notable phenomenon known as frame dragging. We perform a dragging coordinate transformation as\cite{45,46,46a} 
\begin{eqnarray}\label{eq13}
\Omega&=&\frac{d\phi}{dt}=-\frac{g_{03}}{g_{33}},\cr
&=&\frac{a(2Mr-Q^{2}+\alpha r^{1-3\omega})}{(r^{2}+a^{2})\Sigma^{2}+a^{2}\sin^{2}\theta(2Mr-Q^{2}+\alpha r^{1-3\omega})}.
\end{eqnarray} 
Then, Eq. (\ref{eq11}) becomes
\begin{eqnarray}\label{eq14}
ds^{2}=-\frac{\Delta_{r}}{r^{2}+a^{2}+a^{2}\sin^{2}\theta(\frac{2Mr-Q^{2}+\alpha r^{1-3\omega}}{\Sigma^{2}})}dt^{2}+\frac{\Sigma^{2}}{\Delta_{r}}dr^{2}+\Sigma^{2}d\theta^{2}.
\end{eqnarray}
The surface gravity of the KNBHQ near the event horizon is given by\cite{47,47a,47b}
\begin{eqnarray}\label{eq15}
\kappa &=& \lim_{g_{00}\to 0} \left( -\frac{1}{2}\sqrt{-\frac{g^{11}}{g_{00}}}\frac{dg_{00}}{dr}\right),\cr &=& \frac{r_{h}-M-\frac{r_{h}^{-3\omega}\alpha (1-3\omega)}{2}}{r_{h}^{2}+a^{2}}.
\end{eqnarray}

Defining the event horizon of the KNBHQ by the relation
\begin{eqnarray}\label{eq16}
r^{2}_{h}-2Mr_{h}+a^{2}+Q^{2}-\alpha r_{h}^{1-3\omega}=0,
\end{eqnarray}
the mass of the KNBHQ can be expressed as
\begin{eqnarray}\label{eq17}
M=\frac{1}{2}\left( r_{h}+\frac{a^{2}}{r_{h}}+\frac{Q^{2}}{r_{h}}-\alpha r_{h}^{-3\omega}\right).
\end{eqnarray}
Here, $ r_{h} $ is the event horizon of the KNBHQ.

\section{EGUP effects on the the thermodynamic quantities of KNBHQ}
Following the first law of black hole mechanics\cite{5} and the definition of thermodynamics, Hawking temperature is expressed by Xiang and Wen\cite{48} as 
\begin{eqnarray}\label{eq18}
T=\left( \frac{dM}{dS}\right) =\frac{dM}{dA}\times\frac{dA}{dS}=\frac{\kappa}{8\pi}\times\frac{dA}{dS},
\end{eqnarray}
where $ A $ and $ S $ are the area and entropy of the black hole.

According to the results derived from GUP by Medved and Vagnas \cite{49}, we get
\begin{eqnarray}\label{eq19}
\frac{dA}{dS}\simeq\frac{(\Delta A)_{min}}{(\Delta S)_{min}}\simeq \frac{\epsilon}{\ln 2} \Delta x \Delta p.
\end{eqnarray}
The calibration factor $ \epsilon $ can be simplified to represent the original form of the temperature of a KNBHQ. Here, we use $ \epsilon=8 \ln 2 $. Assuming the uncertainty in the position is in the order of the diameter of the black hole, $ \Delta x\simeq 2 r_{h} $ and utilizing the momentum uncertainty $ \Delta p $ as described in Eq. (\ref{eq9}), we derive the modified Hawking temperature influenced by EGUP as
\begin{eqnarray}\label{eq20}
T_{EGUP}=\frac{4r^{2}_{h}\left( r_{h}-M-\frac{r_{h}^{-3\omega}\alpha (1-3\omega)}{2}\right)}{\pi\beta\left( r_{h}^{2}+a^{2}\right)}\Big[1-\sqrt{1-\beta\left( \frac{1}{4 r^{2}_{h}}+\frac{\eta}{L_{\ast}^{2}}\right)}\Big].
\end{eqnarray}
When $ \eta=0 $, in Eq. (\ref{eq20}) we get the Hawking temperature of the KNBHQ modified by GUP as
\begin{eqnarray}\label{eq21}
T_{GUP}=\frac{4r^{2}_{h}\left( r_{h}-M-\frac{r_{h}^{-3\omega}\alpha (1-3\omega)}{2}\right)}{\pi\beta\left( r_{h}^{2}+a^{2}\right)}\Big[1-\sqrt{1-\frac{\beta}{4r^{2}_{h}}}\Big].
\end{eqnarray}
When $ \beta=0 $, in Eq. (\ref{eq20}) we get EUP modified Hawking temperature of KNBHQ as
\begin{eqnarray}\label{eq22}
T_{EUP}=\frac{4r^{2}_{h}\left( r_{h}-M-\frac{r_{h}^{-3\omega}\alpha (1-3\omega)}{2}\right)}{2\pi\left( r_{h}^{2}+a^{2}\right)}\left( \frac{1}{4r^{2}_{h}}+\frac{\eta}{L^{2}_{\ast}}\right).
\end{eqnarray}
When $ \beta=0 $ and $ \eta=0 $, in Eq. (\ref{eq20}) we get HUP modified Hawking temperature of KNBHQ as
\begin{eqnarray}\label{eq23}
T_{HUP}=\frac{ r_{h}-M-\frac{r_{h}^{-3\omega}\alpha (1-3\omega)}{2}}{2\pi\left( r_{h}^{2}+a^{2}\right)}.
\end{eqnarray}
\begin{figure}[h!]
    \centering
    \begin{subfigure}{0.49\textwidth}
        \centering
        \includegraphics[width=\linewidth]{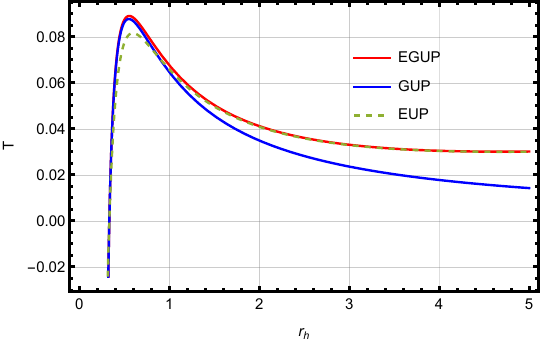} 
        \caption{}
    \end{subfigure}
    \hfill
    \begin{subfigure}{0.49\textwidth}
        \centering
        \includegraphics[width=\linewidth]{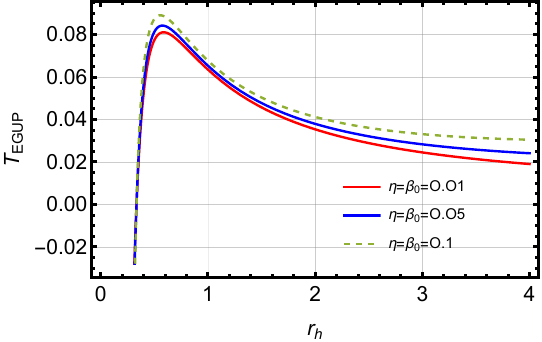} 
        \caption{}
    \end{subfigure}
    \vskip\baselineskip
    \begin{subfigure}{0.49\textwidth}
        \centering
        \includegraphics[width=\linewidth]{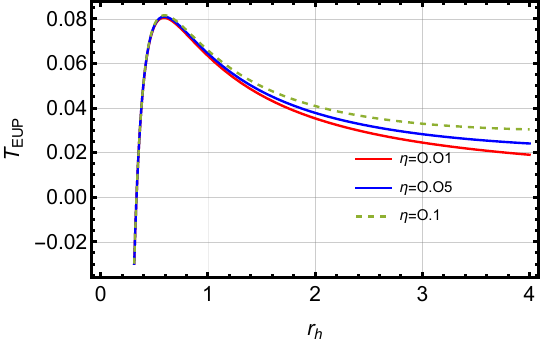} 
        \caption{}
    \end{subfigure}
    \hfill
    \begin{subfigure}{0.49\textwidth}
        \centering
        \includegraphics[width=\linewidth]{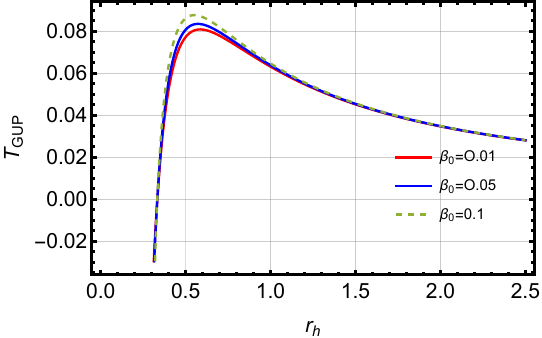} 
        \caption{}
    \end{subfigure}
    \caption{Hawking temperature as a function of $ r_{h} $ for $a = 0.1$, $L_{\ast} = 3$, $Q=0.3$, $ \omega=-\frac{1}{3}  $ and $ \alpha=0.1$. (a) Comparison of the Hawking temperature of EGUP, GUP and EUP. (b) Corrected Hawking temperature of EGUP for different $ \eta $ and $ \beta_{0} $. (c) Corrected Hawking temperature of EUP for different $ \eta $. (d) Corrected Hawking temperature of GUP for different $ \beta_{0} $.}
\label{fig1}
\end{figure}

\begin{figure}[h!]
    \centering
    \begin{subfigure}{0.49\textwidth}
        \centering
        \includegraphics[width=\linewidth]{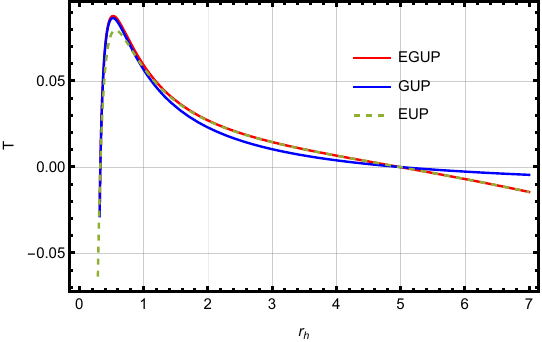} 
        \caption{}
    \end{subfigure}
    \hfill
    \begin{subfigure}{0.49\textwidth}
        \centering
        \includegraphics[width=\linewidth]{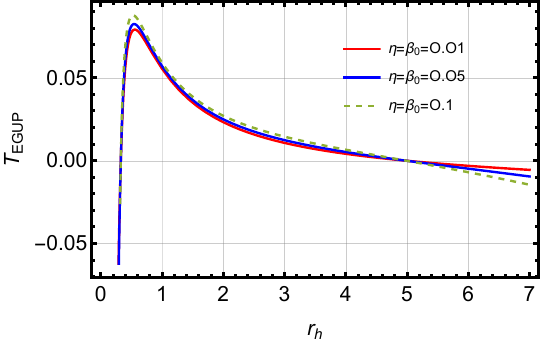} 
        \caption{}
    \end{subfigure}
    \vskip\baselineskip
    \begin{subfigure}{0.49\textwidth}
        \centering
        \includegraphics[width=\linewidth]{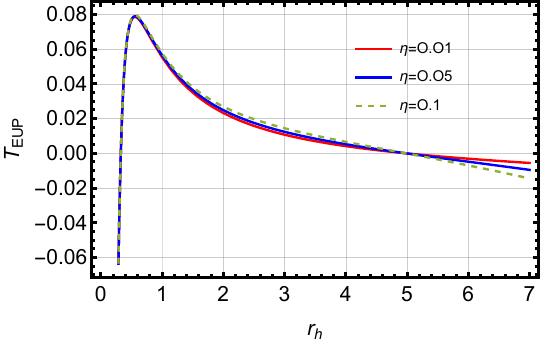} 
        \caption{}
    \end{subfigure}
    \hfill
    \begin{subfigure}{0.49\textwidth}
        \centering
        \includegraphics[width=\linewidth]{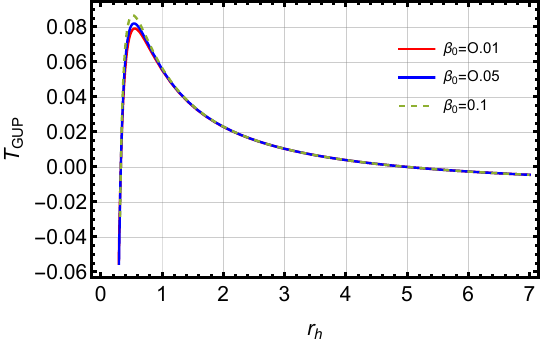} 
        \caption{}
    \end{subfigure}
    \caption{Hawking temperature as a function of $ r_{h} $ for $a = 0.1$, $L_{\ast} = 3$, $Q=0.3$, $ \omega=-\frac{2}{3}  $ and $ \alpha=0.1$. (a) Comparison of the Hawking temperature of EGUP, GUP and EUP. (b) Corrected Hawking temperature of EGUP for different $ \eta $ and $ \beta_{0} $. (c) Corrected Hawking temperature of EUP for different $ \eta $. (d) Corrected Hawking temperature of GUP for different $ \beta_{0} $.}
\label{fig2}
\end{figure}

Considering that the Hawking temperature is positive and has a real value, we need to put some constraints on Eq. (\ref{eq20}). The last two terms of Eq. (\ref{eq20}) must yield positive values, and the expression within the square root must also be positive. We analyze two scenarios for the parameters of the quintessence field, specifically $ \omega=-\frac{1}{3} $ and $ \omega=-\frac{2}{3} $. As a result, following constraints on the event horizon radius are established as
\begin{eqnarray}\label{eq24}
\frac{r^{2}_{h}(1-\alpha)-a^{2}-Q^{2}}{2r_{h}(r^{2}_{h}+a^{2})}\geq 0,\ for\ \omega=-\frac{1}{3}
\end{eqnarray}
and
\begin{eqnarray}\label{eq25}
\frac{r^{2}_{h}(1-2r_{h}\alpha)-a^{2}-Q^{2}}{2r_{h}(r^{2}_{h}+a^{2})}\geq 0,\ for\ \omega=-\frac{2}{3}.
\end{eqnarray}
Eq. (\ref{eq24}) establishes a minimum limit on the horizon $ r_{h} $ as
\begin{eqnarray}\label{eq26}
r_{h}\geq\sqrt{\frac{a^{2}+Q^{2}}{(1-\alpha)}},\ 0\leq\alpha \leq1.
\end{eqnarray}
Clearly, Eq. (\ref{eq25}) has three roots: one real root and a pair of imaginary roots. In this context, we present only the significant real root.
\begin{eqnarray}\label{eq27}
r_{h}=\frac{1}{6\alpha}-\frac{1}{3\times 2^{\frac{2}{3}}\alpha(\zeta+\sqrt{-4+\zeta^{2}})^{\frac{1}{3}}}-\frac{(\zeta +\sqrt{-4+\zeta^{2}})^{\frac{1}{3}}}{6\times 2^{\frac{1}{3}}\alpha},
\end{eqnarray}
where $ \zeta=-2+108a^{2}\alpha^{2}+108Q^{2}\alpha^{2} $.

Considering the second constraint containing EGUP parameters, we have
\begin{eqnarray}\label{eq28}
1-\beta\left(\frac{1}{4r^{2}_{h}}+\frac{\eta}{L^{2}_{\ast}} \right) \geq 0,
\end{eqnarray}
this constraint gives the minimum value of the horizon as
\begin{eqnarray}\label{eq29}
r_{h(EGUP)}\geq\frac{L_{\ast}}{2}\sqrt{\frac{\beta}{L^{2}_{\ast}-\eta\beta}}.
\end{eqnarray}
In case of GUP, Eq. (\ref{eq29}) reduces to
\begin{eqnarray}\label{eq30}
r_{h(GUP)}\geq\frac{\sqrt{\beta}}{2}.
\end{eqnarray}
However, in the case of EUP and HUP, the constraint disappears, resulting in $ r_{h}\geq 0 $. The requirement for the black hole's event horizon to be non-zero, as stated in Eqs. (\ref{eq29}) and (\ref{eq30}), indicates that the black hole has not entirely evaporated due to thermal radiation.
\begin{figure}[h!]
    \centering
    \begin{subfigure}{0.49\textwidth}
        \centering
        \includegraphics[width=\linewidth]{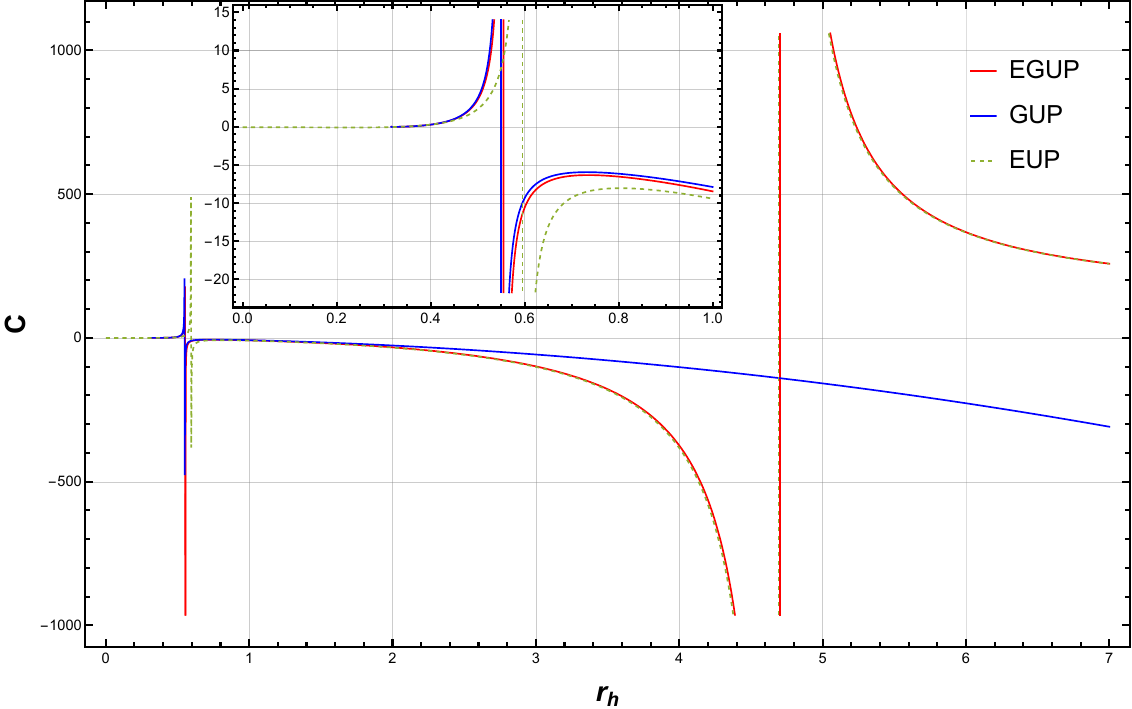} 
        \caption{}
    \end{subfigure}
    \hfill
    \begin{subfigure}{0.49\textwidth}
        \centering
        \includegraphics[width=\linewidth]{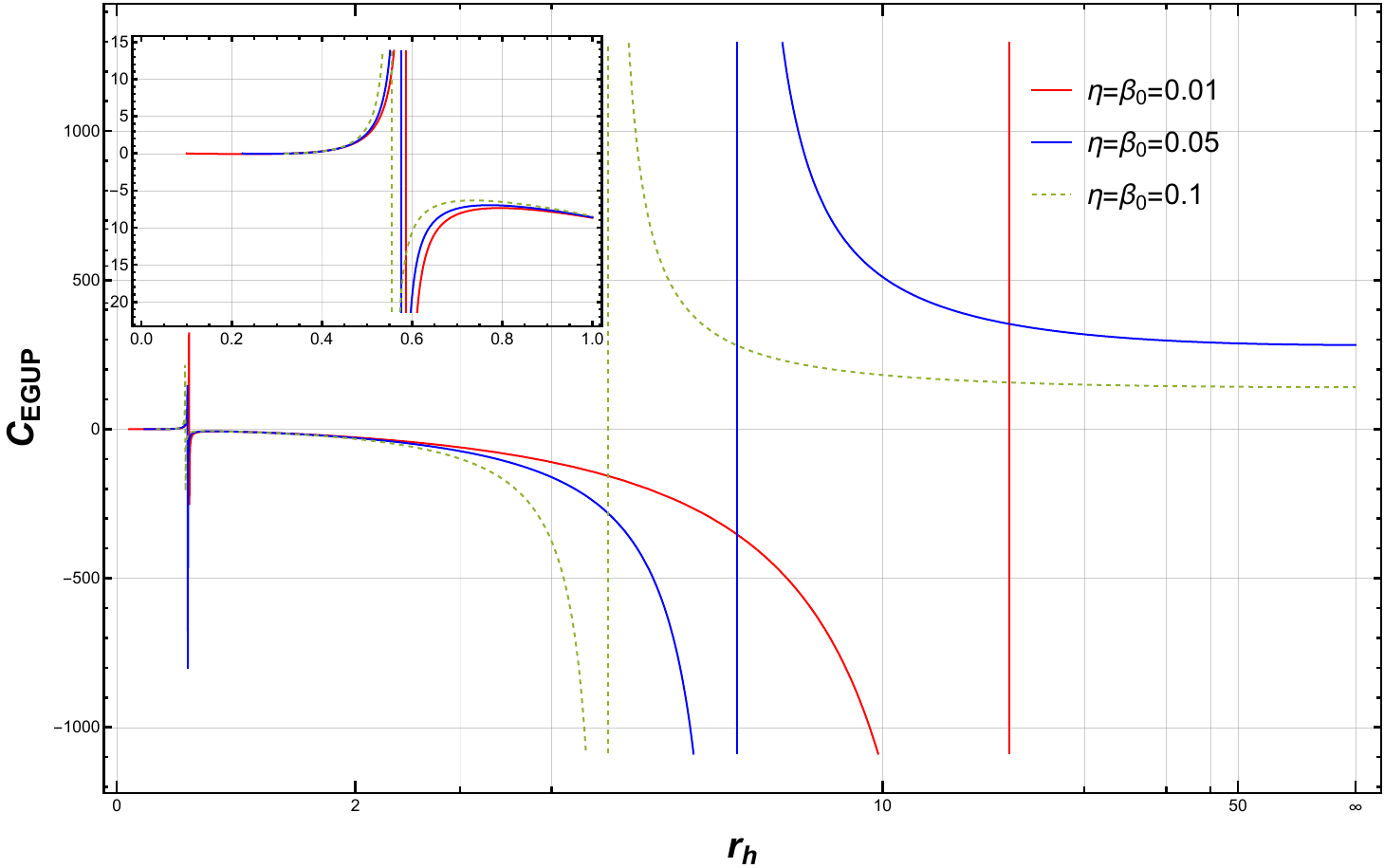} 
        \caption{}
        \label{fig3b}
    \end{subfigure}
    \vskip\baselineskip
    \begin{subfigure}{0.49\textwidth}
        \centering
        \includegraphics[width=\linewidth]{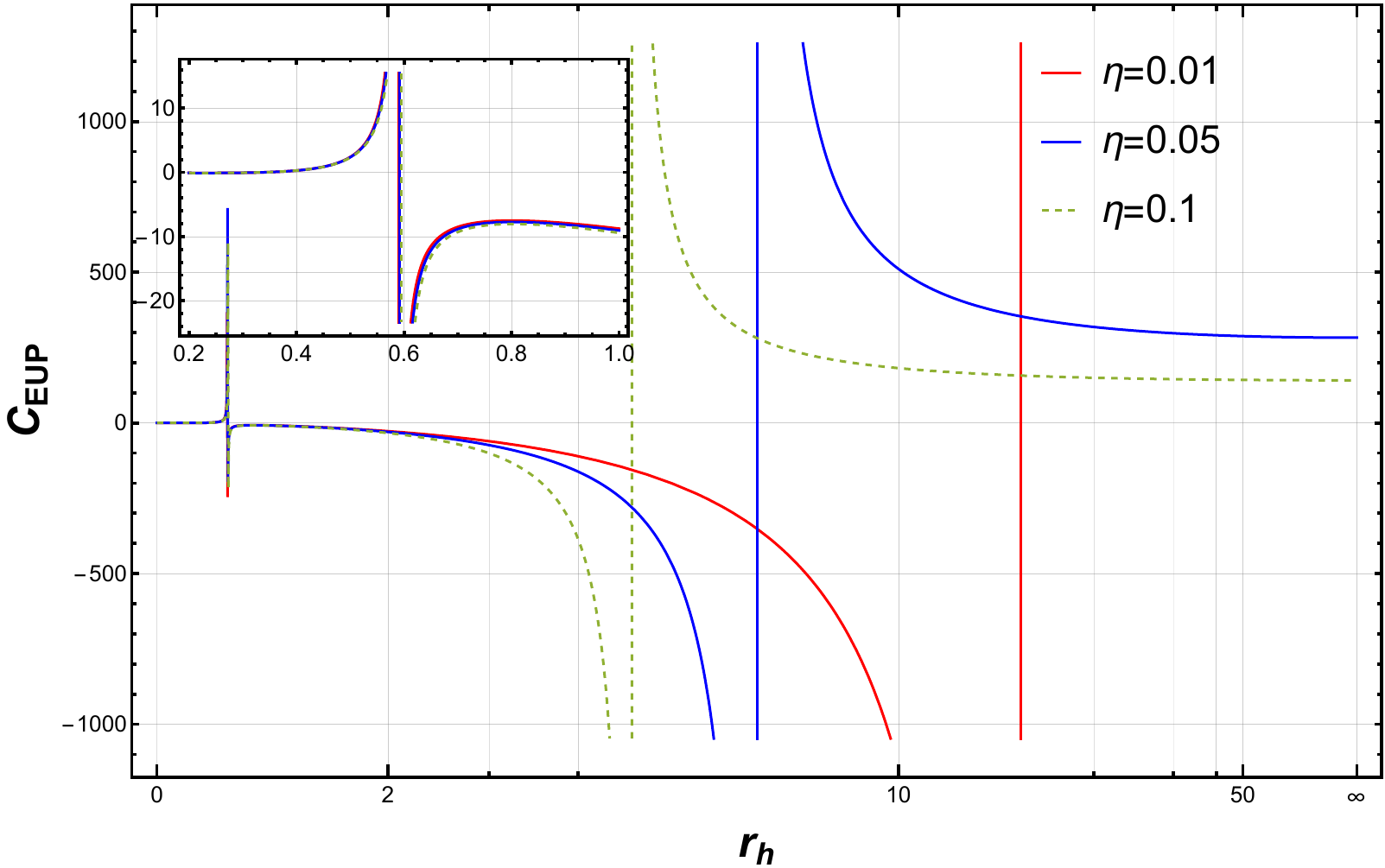} 
        \caption{}
        \label{fig3c}
    \end{subfigure}
    \hfill
    \begin{subfigure}{0.49\textwidth}
        \centering
        \includegraphics[width=\linewidth]{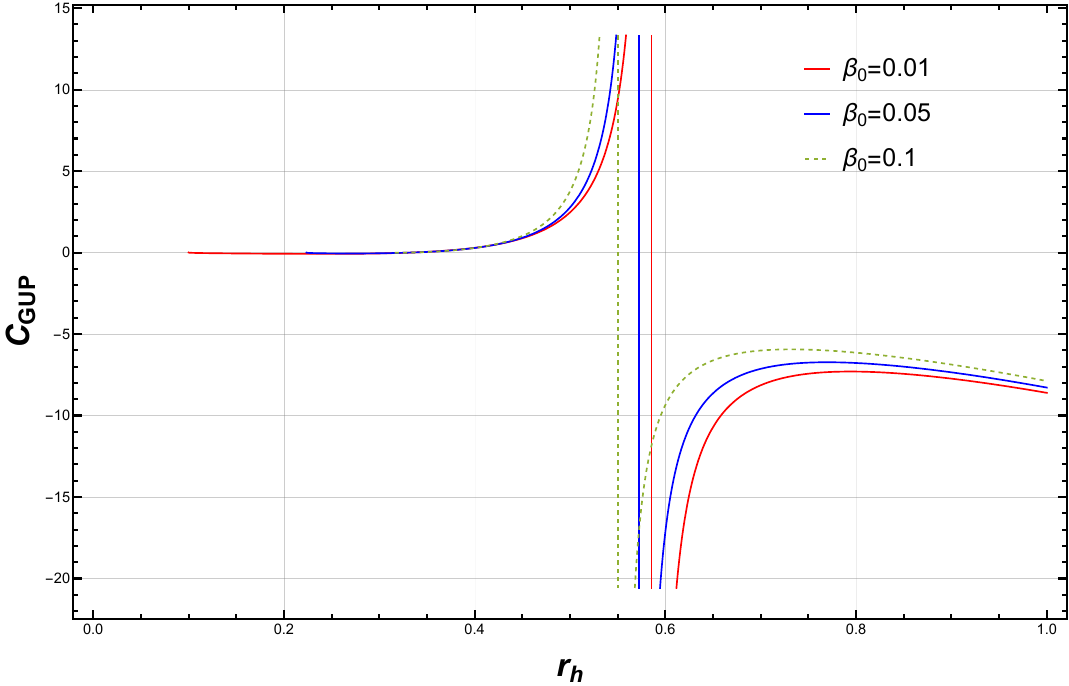} 
        \caption{}
    \end{subfigure}
    \caption{Heat capacity as a function of $ r_{h} $ for $a = 0.1$, $L_{\ast} = 3$, $Q=0.3$, $ \omega=-\frac{1}{3}  $ and $ \alpha=0.1$. (a) Comparison of of EGUP, GUP and EUP heat capacities. (b) Heat capacity of EGUP for different $ \eta $ and $ \beta_{0} $. (c) Heat capacity of EUP for different $ \eta $. (d) Heat capacity of GUP for different $ \beta_{0} $. }
\label{fig3}
\end{figure}

\begin{figure}[h!]
    \centering
    \begin{subfigure}{0.49\textwidth}
        \centering
        \includegraphics[width=\linewidth]{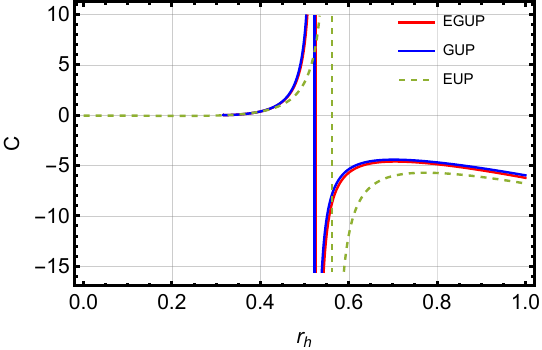} 
        \caption{}
    \end{subfigure}
    \hfill
    \begin{subfigure}{0.49\textwidth}
        \centering
        \includegraphics[width=\linewidth]{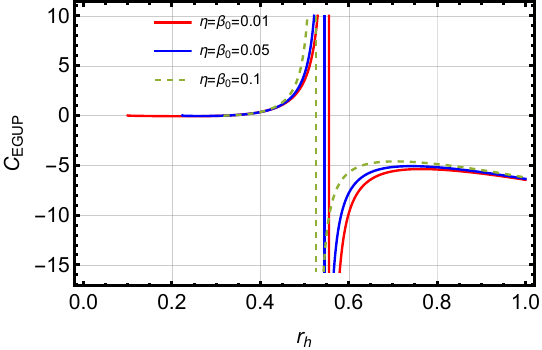} 
        \caption{}
    \end{subfigure}
    \vskip\baselineskip
    \begin{subfigure}{0.49\textwidth}
        \centering
        \includegraphics[width=\linewidth]{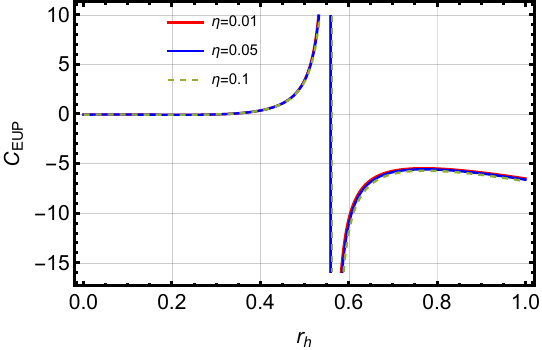} 
        \caption{}
    \end{subfigure}
    \hfill
    \begin{subfigure}{0.49\textwidth}
        \centering
        \includegraphics[width=\linewidth]{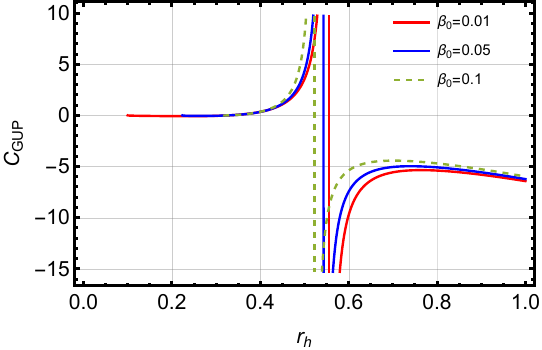} 
        \caption{}
    \end{subfigure}
    \caption{Heat capacity as a function of $ r_{h} $ for $a = 0.1$, $L_{\ast} = 3$, $Q=0.3$, $ \omega=-\frac{2}{3}  $ and $ \alpha=0.1$. (a) Comparison of of EGUP, GUP and EUP heat capacities. (b) Heat capacity of EGUP for different $ \eta $ and $ \beta_{0} $. (c) Heat capacity of EUP for different $ \eta $. (d) Heat capacity of GUP for different $ \beta_{0} $.}
\label{fig4}
\end{figure}

The modifications of Hawking temperature as influenced by the effects of EGUP, GUP and EUP are illustrated in Figs. \ref{fig1} and \ref{fig2} for the quintessence field parameter $ \omega=-\frac{1}{3} $ and $ \omega=-\frac{2}{3} $, respectively. In this analysis, the parameters utilized include $a = 0.1$, $L_{\ast} = 3$, $Q=0.3$ and $\alpha=0.1$. According to the graph depicted in Figs. \ref{fig1} and \ref{fig2}, the modified temperatures are positive for $ r_{h} \geq 0.330 $ when $ \omega = -\frac{1}{3} $. In contrast, for $ \omega = -\frac{2}{3} $, the modified temperatures are positive and yield real values within the interval $ 0.320 < r_{h} < 5 $.

Black holes exhibit thermodynamic stability when their heat capacity is positive. They are considered unstable when their heat capacity is negative. The heat capacity of a black hole may be either positive or negative, contingent upon the size of the black hole and various other factors.

The heat capacity of the KNBHQ, modified by EGUP in the presence of quintessence, can be expressed as follows
\begin{eqnarray}\label{eq31}
C_{EGUP}&=&\frac{dM}{dT_{EGUP}}=\frac{\partial M}{\partial r_{h}}\times \frac{\partial r_{h}}{\partial T_{EGUP}}\cr &=& -\pi r^{3\omega}_{h}(r^{2}_{h}+a^{2})^{2}\lbrace a^{2}+Q^{2}-r_{h}(r_{h}+3r^{-3\omega}_{h}\alpha\omega)\rbrace \cr&& \times \Big[2\Big[ \frac{(r^{2}_{h}+a^{2})\lbrace r^{3\omega}_{h}(a^{2}+Q^{2}-r^{2}_{h})-3r_{h}\alpha\omega\rbrace}{2\sqrt{1-\beta\left( \frac{1}{4 r^{2}_{h}}+\frac{\eta}{L_{\ast}^{2}}\right)}}\cr&&-\frac{2r^{2}_{h}a^{2}}{\beta}\left\lbrace  -2+2\sqrt{1-\beta\left( \frac{1}{4 r^{2}_{h}}+\frac{\eta}{L_{\ast}^{2}}\right)}\right\rbrace-\frac{r^{2}_{h}(r^{2}_{h}+a^{2})}{\beta}\cr&& \times\left\lbrace  -2+2\sqrt{1-\beta\left( \frac{1}{4 r^{2}_{h}}+\frac{\eta}{L_{\ast}^{2}}\right)}\right\rbrace \lbrace r^{3\omega}_{h}(a^{2}+Q^{2}+r^{2}_{h})\cr&&-9r_{h}\alpha\omega^{2}\rbrace\Big] \Big]^{-1}. 
\end{eqnarray}
Similarly, the heat capacities modified by GUP and EUP are given by
\begin{eqnarray}\label{eq32}
C_{GUP}&=&\pi\beta(r^{2}_{h}+a^{2})\sqrt{4-\frac{\beta}{r^{2}_{h}}}\left( r^{2-3\omega}_{h}-a^{2}r^{3\omega}_{h}-Q^{2}r^{3\omega}_{h}+3r_{h}\alpha\omega\right)\cr&& \times \Big[2r_{h}\Big[2a^{4}r^{1-3\omega}_{h}\left(2-\sqrt{4-\frac{\beta}{r^{2}_{h}}} \right)+a^{2}\lbrace4\beta r^{1-3\omega}_{h}\cr&& +2Q^{2}r^{1-3\omega}_{h}\left(2-\sqrt{4-\frac{\beta}{r^{2}_{h}}} \right)-8r^{3-3\omega}_{h}\left(2-\sqrt{4-\frac{\beta}{r^{2}_{h}}} \right)\cr&& +3\beta\alpha(1-3\omega)\omega +6r^{2}_{h}\alpha\left(2-\sqrt{4-\frac{\beta}{r^{2}_{h}}} \right)\omega(3\omega-2)\rbrace \cr&&+r_{h}\lbrace -2r^{4-3\omega}_{h}\left(2-\sqrt{4-\frac{\beta}{r^{2}_{h}}} \right)+2Q^{2}r^{3\omega}_{h}\cr&& \times\lbrace \beta-r^{2}_{h}\left(2-\sqrt{4-\frac{\beta}{r^{2}_{h}}} \right)\rbrace +18r^{3}_{h}\alpha\left(2-\sqrt{4-\frac{\beta}{r^{2}_{h}}} \right)\omega^{2}\cr&& -3\beta r_{h}\alpha\omega(1+3\omega)\rbrace  \Big]\Big]^{-1}
\end{eqnarray}
and
\begin{eqnarray}\label{eq33}
C_{EUP}&=&-2L^{2}_{\ast}\pi r^{3\omega}_{h}\left(r^{2}_{h}+a^{2} \right)^{2}\left\lbrace a^{2}+Q^{2}-r_{h}(r_{h}+3r^{-3\omega}_{h}\alpha\omega)\right\rbrace \cr&&\times \Big[r^{3\omega}_{h}\lbrace L^{2}_{\ast}\lbrace a^{4}+3Q^{2}r^{2}_{h}-r^{4}_{h}+a^{2}(Q^{2}+4r^{2}_{h})\rbrace \cr&& +4r^{2}_{h}\lbrace -a^{2}(a^{2}+Q^{2})+(4a^{2}+Q^{2})r^{2}_{h}+r^{4}_{h} \rbrace \eta \rbrace +3r_{h}\alpha\omega \cr&& \times \lbrace -2r^{2}_{h}(L^{2}_{\ast}-4a^{2}\eta)-3(a^{2}+r^{2})(L^{2}_{\ast}+4r^{2}_{h}\eta)\omega \rbrace\Big]^{-1}. 
\end{eqnarray}

The variations in the specific heat capacity of the KNBHQ, influenced by EGUP, EUP and GUP with respect to the event horizon radius $ r_{h} $, are presented in Figs. \ref{fig3} and \ref{fig4} for the quintessence field parameters $ \omega=-\frac{1}{3} $ and $ \omega=-\frac{2}{3} $, respectively. In this analysis, the parameters utilized include $a = 0.1, L_{\divideontimes} = 3, Q=0.3,$ and $ \alpha=0.1$. From Fig. \ref{fig3}, for the case of $\omega = -\tfrac{1}{3}$, it is observed that the heat capacity for the EGUP and EUP cases exhibits two distinct divergence points, indicating the presence of two second-order phase transitions. Between these divergence points, the heat capacity changes sign, dividing the system into three thermodynamic regimes: a small black hole phase and a large black hole phase, both characterized by positive heat capacity and hence thermodynamic stability, and an intermediate phase with negative heat capacity, indicating instability. In contrast, for the GUP case, the heat capacity displays only a single divergence point, corresponding to a single second-order phase transition. The phase transition points are significantly influenced by the correction parameters in all three cases. For EGUP,  it is observed that the first phase transition point is only weakly affected by variations in the parameters $\eta$ and $\beta_0$, whereas the second phase transition point shifts considerably, indicating a strong sensitivity in the large black hole regime. A similar behavior is found in the EUP case.
Furthermore, for $\omega = -\tfrac{2}{3}$, only one divergence point is observed for all three cases, implying the existence of a single phase transition in this regime. The phase transition points for different values of the correction parameters for the EGUP, EUP and GUP cases are presented in Table \ref{tabheat}. In the EGUP framework, for both $\omega=-1/3$ and $\omega=-2/3$, the phase transition points occur at lower event horizon as the parameters $\eta$ and $\beta$ increase. Similarly, in the GUP case, the phase transition points also shift toward lower event horizon with increasing $\beta$ for both choices of $\omega$.
In contrast, the EUP case shows that the first transition point increases gradually with increasing $\eta$ for both $\omega=-1/3$ and $\omega=-2/3$. However, for $\omega=-1/3$, the second transition point decreases significantly as $\eta$ increases. Therefore, while EGUP and GUP corrections generally suppress the transition points, the EUP correction enhances the first phase transition point but reduces the second one.

\begin{table*}[t]
\centering
\renewcommand{\arraystretch}{1.25}
\setlength{\tabcolsep}{8pt}
\begin{tabular}{|c|c|c|c|c|}
\hline
Model & $\omega$ & Parameter & First  phase & Second phase \\
 	&	&	&  Transition Point & Transition Point\\
\hline

\multirow{6}{*}{EGUP}
& \multirow{3}{*}{$-1/3$}
& $\eta=\beta=0.01$ & 0.585862 & 14.9734 \\ \cline{3-5}
& & $\eta=\beta=0.05$ & 0.575164 & 6.66185 \\ \cline{3-5}
& & $\eta=\beta=0.10$ & 0.554304 & 4.70254 \\ \cline{2-5}
& \multirow{3}{*}{$-2/3$}
& $\eta=\beta=0.01$ & 0.555161 & -- \\ \cline{3-5}
& & $\eta=\beta=0.05$ & 0.544630 & -- \\ \cline{3-5}
& & $\eta=\beta=0.10$ & 0.525325 & -- \\
\hline

\multirow{6}{*}{EUP}
& \multirow{3}{*}{$-1/3$}
& $\eta=0.01$ & 0.585841 & 14.9588 \\ \cline{3-5}
& & $\eta=0.05$ & 0.592002 & 6.67067 \\ \cline{3-5}
& & $\eta=0.10$ & 0.595080 & 4.69689 \\ \cline{2-5}
& \multirow{3}{*}{$-2/3$}
& $\eta=0.01$ & 0.557770 & -- \\ \cline{3-5}
& & $\eta=0.05$ & 0.559534 & -- \\ \cline{3-5}
& & $\eta=0.10$ & 0.561887 & -- \\
\hline

\multirow{6}{*}{GUP}
& \multirow{3}{*}{$-1/3$}
& $\beta=0.01$ & 0.585522 & -- \\ \cline{3-5}
& & $\beta=0.05$ & 0.571844 & -- \\ \cline{3-5}
& & $\beta=0.10$ & 0.549841 & -- \\ \cline{2-5}
& \multirow{3}{*}{$-2/3$}
& $\beta=0.01$ & 0.555242 & -- \\ \cline{3-5}
& & $\beta=0.05$ & 0.543004 & -- \\ \cline{3-5}
& & $\beta=0.10$ & 0.522423 & -- \\
\hline

\end{tabular}
\caption{Comparison of phase transition points of heat capacity for EGUP, EUP and GUP models for different values of $\omega$.}
\label{tabheat}
\end{table*}

Vanishing of the heat capacity implies the formation of black hole remnant\cite{50,51}. The mass of the black hole at which the heat capacity vanishes gives the remnant mass. In the case of EGUP we examine that a black hole remnant occurs for $ r_{h}=\frac{L_{\ast}}{2}\sqrt{\frac{\beta}{L^{2}_{\ast}-\eta\beta}} $. Accordingly, we determine the remnant temperature as 

\begin{eqnarray}\label{eq34}
T_{rem(EGUP)}&=&\frac{L^{2}_{\ast}\left( L_{\ast}\sqrt{\frac{\beta}{L^{2}_{\ast}-\eta\beta}}\right)^{-1-3\omega} }{\pi(\beta \eta - L^{2}_{\ast})\left\lbrace \beta L^{2}_{\ast} +4a^{2}(L^{2}_{\ast}-\beta \eta) \right\rbrace }\Big[ \left(L_{\ast}\sqrt{\frac{\beta}{L^{2}_{\ast}-\eta\beta}} \right)^{3\omega} \cr&& \times \left\lbrace L^{2}_{\ast}\left( 4(a^{2}+Q^{2})-\beta\right) -4\beta \eta(a^{2}+Q^{2}) \right\rbrace \cr&& -3\times 2^{1+3\omega}L_{\ast}\alpha\sqrt{\frac{\beta}{L^{2}_{\ast}-\eta\beta}}(L^{2}_{\ast}-\beta \eta)\omega \Big]
\end{eqnarray}
and the remnant mass is
\begin{eqnarray}\label{eq35}
M_{rem(EGUP)}&=&\frac{1}{2}\Big[\frac{\sqrt{\frac{\beta}{L^{2}_{\ast}-\eta\beta}}\left\lbrace L^{2}_{\ast}(4Q^{2}+\beta)-4Q^{2}\beta\eta +4a^{2}(L^{2}_{\ast}-\beta \eta) \right\rbrace }{2L_{\ast}\beta}\cr&& -8^{\omega}\alpha\left(\sqrt{\frac{\beta}{L^{2}_{\ast}-\eta\beta}} \right)^{-3\omega} \Big].
\end{eqnarray}

In the case of GUP, heat capacity is found to vanish at $ r_{h}=\frac{\sqrt{\beta}}{2} $. Accordingly, the corresponding remnant temperature is given by
\begin{eqnarray}\label{eq36}
T_{rem(GUP)}=\frac{3\times 2^{1+3\omega}\alpha\omega \beta^{\frac{1-3\omega}{2}}+\beta -4(a^{2}+Q^{2})}{\pi \sqrt{\beta}(4a^{2}+\beta)}
\end{eqnarray}
and the remnant mass is

\begin{eqnarray}\label{eq37}
M_{rem(GUP)}=\frac{4(a^{2}+Q^{2})+\beta- 2^{1+3\omega}\alpha \beta^{\frac{1-3\omega}{2}}}{4\sqrt{\beta}}.
\end{eqnarray}
In the EUP scenario, we observe that there is no remnant temperature and mass that is dependent on $ \eta $.

\subsection{Gibbs free energy}
\begin{figure}[b!]
    \centering
    \begin{subfigure}{0.45\textwidth}
        \includegraphics[width=\linewidth]{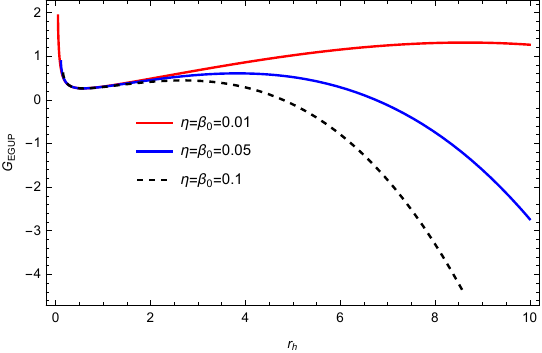}
        \caption{}
    \end{subfigure}
    \hfill
    \begin{subfigure}{0.45\textwidth}
        \includegraphics[width=\linewidth]{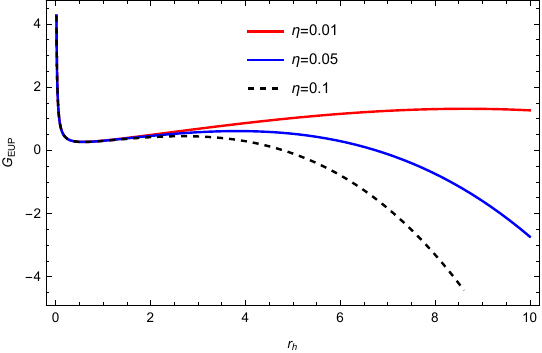}
        \caption{}
    \end{subfigure}
    
    \begin{subfigure}{0.45\textwidth}
        \includegraphics[width=\linewidth]{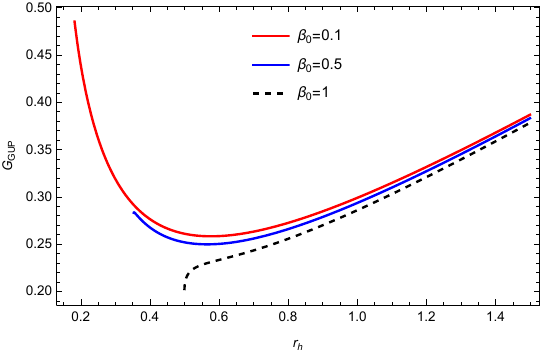}
        \caption{}
    \end{subfigure}
    
    \caption{Gibbs free energy as a function of $ r_{h} $ for $ a=0.1 $, $ Q=0.3 $, $ \alpha=0.1 $, $ L=3 $, and $ \omega=-\frac{1}{3} $. (a)  EGUP corrected Gibbs free energy for different $ \eta $ and $ \beta_{0} $. (b) EUP corrected Gibbs free energy for different $ \eta $. (c) GUP corrected Gibbs free energy for different $ \beta_{0} $.}
\label{fig7}
\end{figure}

\begin{figure}[h!]
    \centering
    \begin{subfigure}{0.45\textwidth}
        \includegraphics[width=\linewidth]{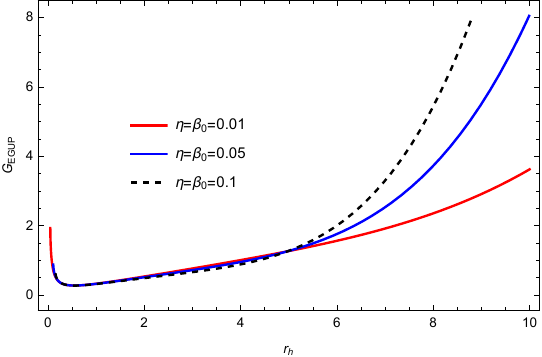}
        \caption{}
    \end{subfigure}
    \hfill
    \begin{subfigure}{0.45\textwidth}
        \includegraphics[width=\linewidth]{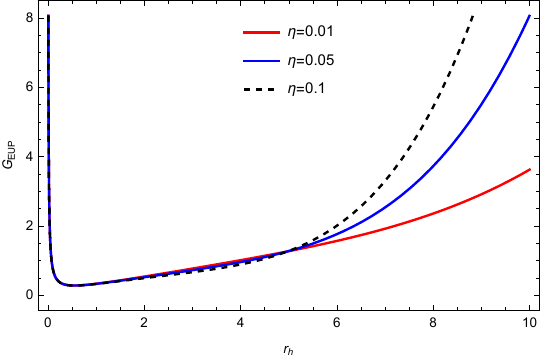}
        \caption{}
    \end{subfigure}
    
    \begin{subfigure}{0.45\textwidth}
        \includegraphics[width=\linewidth]{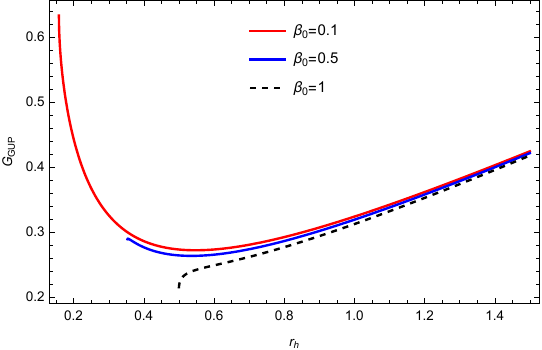}
        \caption{}
    \end{subfigure}
    
    \caption{Gibbs free energy as a function of $ r_{h} $ for $ a=0.1 $, $ Q=0.3 $, $ \alpha=0.1 $, $ L=3 $, and $ \omega=-\frac{2}{3} $. (a)  EGUP corrected Gibbs free energy for different $ \eta $ and $ \beta_{0} $. (b) EUP corrected Gibbs free energy for different $ \eta $. (c) GUP corrected Gibbs free energy for different $ \beta_{0} $.}
\label{fig8}
\end{figure}

\begin{figure}[h!]
    \centering
    \begin{subfigure}{0.45\textwidth}
        \includegraphics[width=\linewidth]{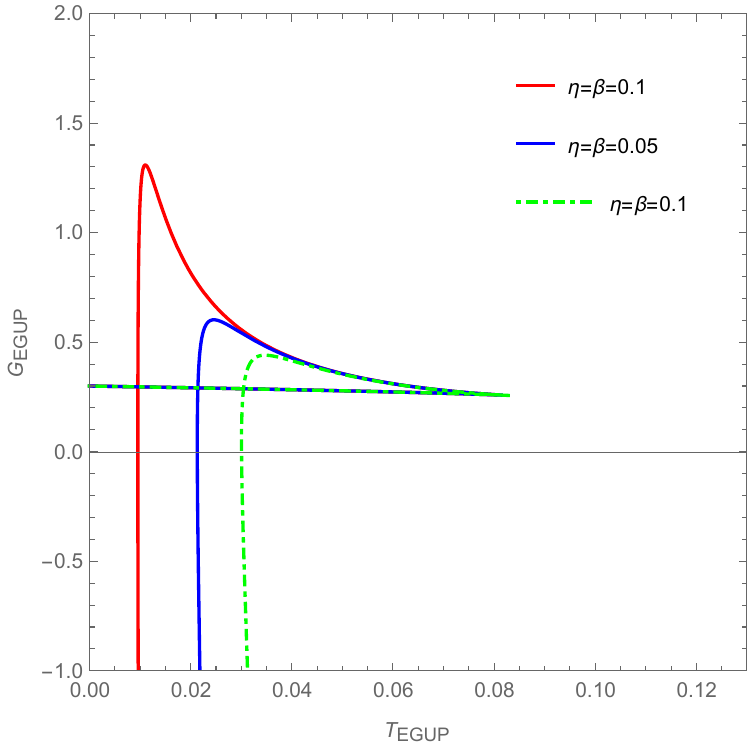}
        \caption{}
    \end{subfigure}
    \hfill
    \begin{subfigure}{0.45\textwidth}
        \includegraphics[width=\linewidth]{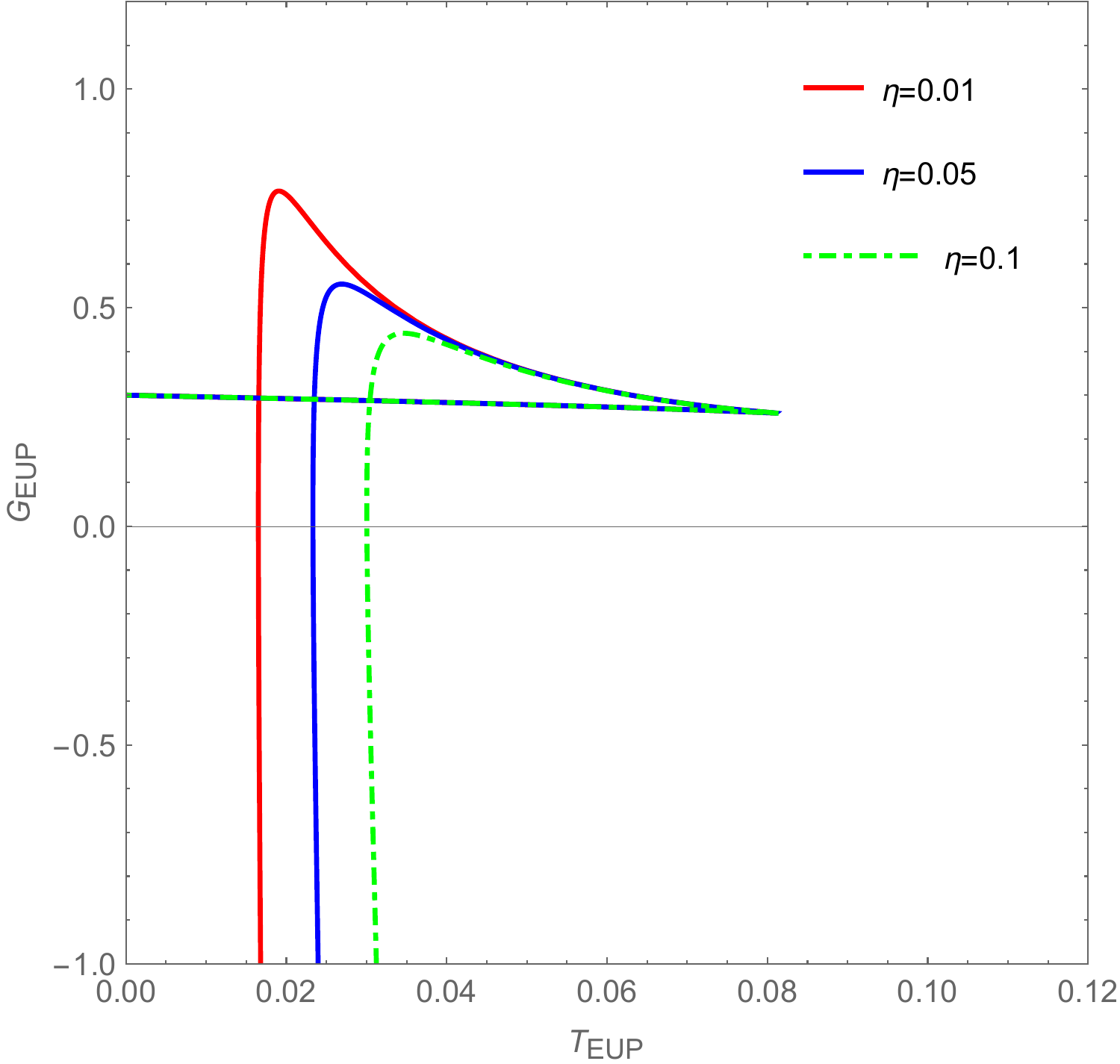}
        \caption{}
    \end{subfigure}
    
    \begin{subfigure}{0.45\textwidth}
        \includegraphics[width=\linewidth]{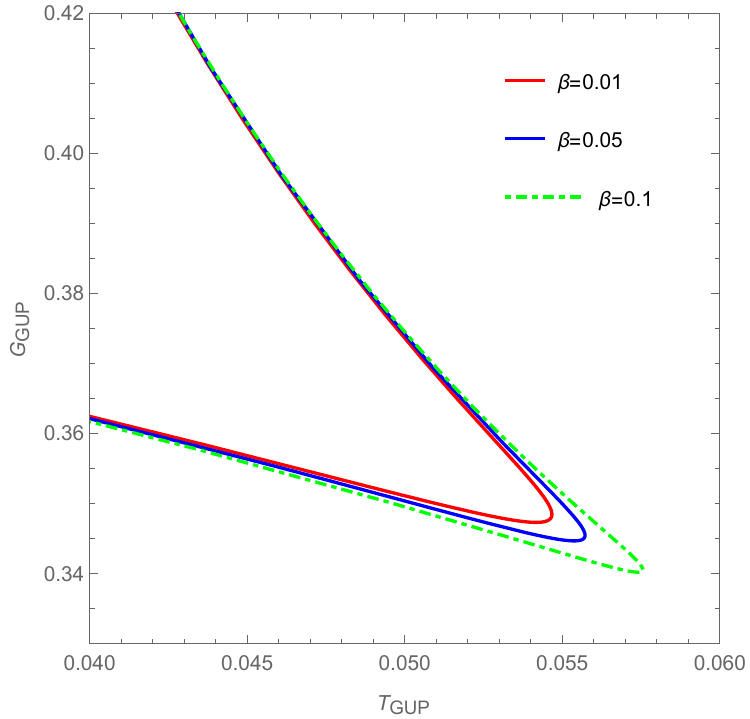}
        \caption{}
    \end{subfigure}
    
    \caption{Gibbs free energy as a function of temperature  for $ a=0.1 $, $ Q=0.3 $, $ \alpha=0.1 $, $ L=3 $, and $ \omega=-\frac{1}{3} $. (a)  EGUP corrected Gibbs free energy for different $ \eta $ and $ \beta_{0} $. (b) EUP corrected Gibbs free energy for different $ \eta $. (c) GUP corrected Gibbs free energy for different $ \beta_{0} $.}
    \label{Gibbs13}
\end{figure}

\begin{figure}[h!]
    \centering
    \begin{subfigure}{0.45\textwidth}
        \includegraphics[width=\linewidth]{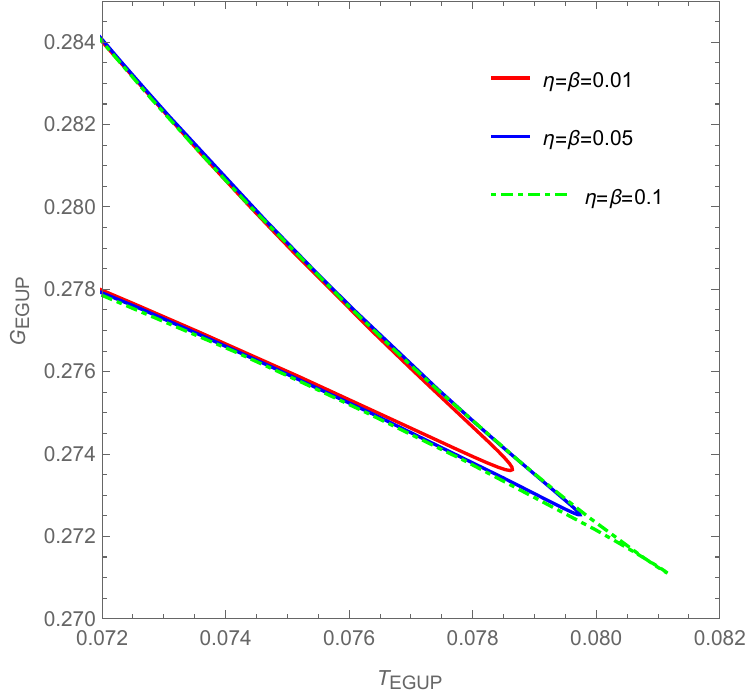}
        \caption{}
    \end{subfigure}
    \hfill
    \begin{subfigure}{0.45\textwidth}
        \includegraphics[width=\linewidth]{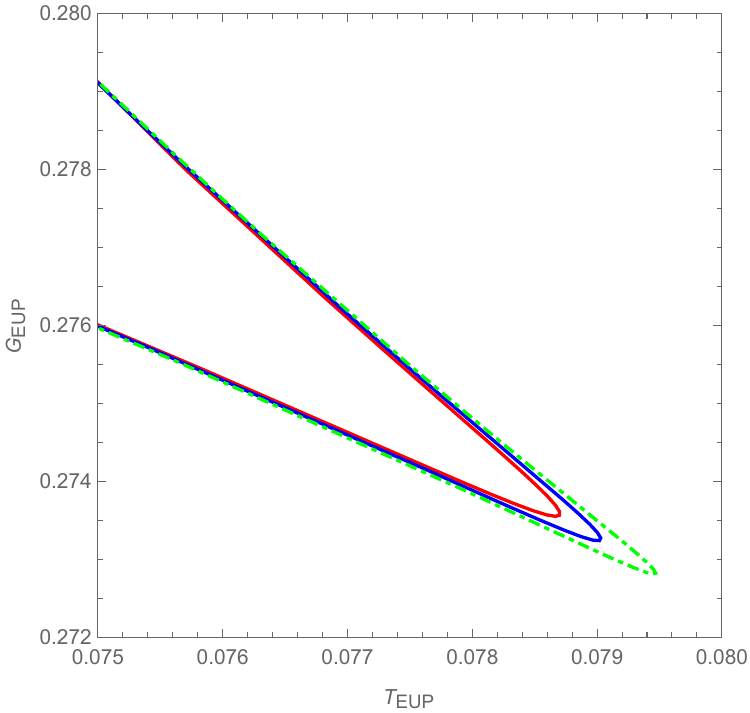}
        \caption{}
    \end{subfigure}
    
    \begin{subfigure}{0.45\textwidth}
        \includegraphics[width=\linewidth]{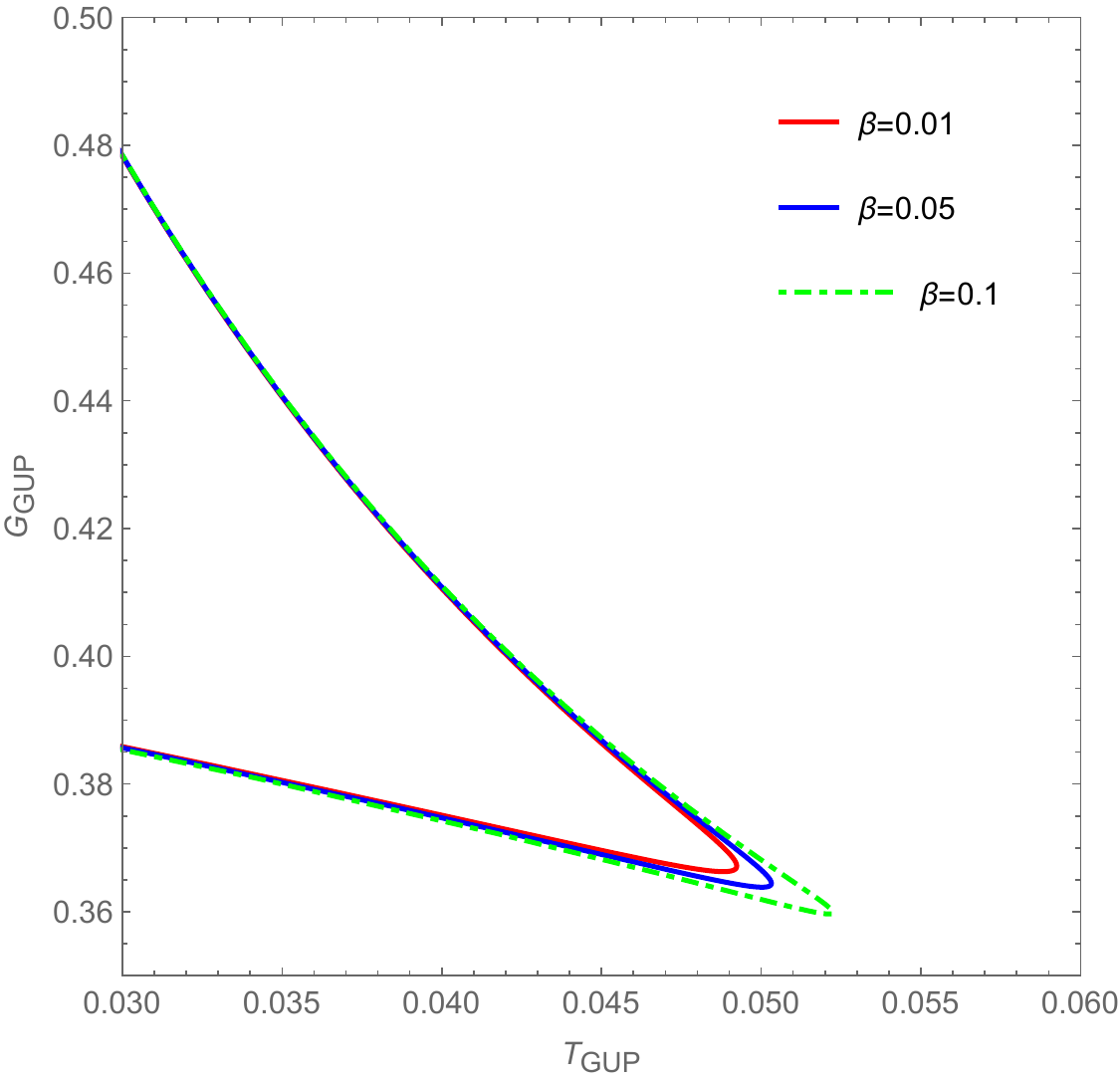}
        \caption{}
    \end{subfigure}
    
    \caption{Gibbs free energy as a function of temperature  for $ a=0.1 $, $ Q=0.3 $, $ \alpha=0.1 $, $ L=3 $, and $ \omega=-\frac{2}{3} $. (a)  EGUP corrected Gibbs free energy for different $ \eta $ and $ \beta_{0} $. (b) EUP corrected Gibbs free energy for different $ \eta $. (c) GUP corrected Gibbs free energy for different $ \beta_{0} $.}
    \label{Gibbs23}
\end{figure}

Gibbs free energy is an important thermodynamic quantity that indicates whether a system at stable temperature and pressure will change spontaneously and which phase is favored thermodynamically. In this study, we show that the EGUP and the quintessence parameter associated with the Kerr-Newman black hole influence the Hawking temperature and entropy. This influence leads to a modification in the Gibbs free energy, resulting to shifted points of phase transition and changed stability regions. By considering \( M \) as enthalpy, we can express Gibbs free energy in canonical ensemble as \cite{57a,57b,57c}
\begin{eqnarray}\label{eq46}
G=M-TS,
\end{eqnarray}
where $ S=\pi (r_{h}^{2}+a^{2}) $.
Using Eq. (\ref{eq20}) into Eq. (\ref{eq46}), we get the EGUP modified Gibbs free energy as
\begin{eqnarray}\label{eq47}
G_{EGUP}&=&\frac{1}{2r_{h}}\Big[a^{2}+Q^{2}+r_{h}\left(r_{h}-r_{h}^{-3\omega}\alpha \right) \Big]-\frac{1}{\beta} \Big[2r_{h}^{1-3\omega}\cr&& \times\left\lbrace  1-\sqrt{1-\beta\left( \frac{1}{4r^{2}_{h}}+\frac{\eta}{L^{2}_{\ast}}\right) } \right\rbrace \lbrace r_{h}^{2+3\omega}-a^{2}r_{h}^{3\omega}-Q^{2}r_{h}^{3\omega}\cr&&+3r_{h}\alpha\omega\rbrace\Big].
\end{eqnarray}
In the same manner, by substituting Eq. (\ref{eq21}), Eq. (\ref{eq22}), Eq. (\ref{eq40}), and Eq. (\ref{eq41}) into Eq. (\ref{eq46}), we derive the Gibbs free energy modified by GUP and EUP as 
\begin{eqnarray}\label{eq48}
G_{GUP}&=&\frac{1}{2r_{h}}\Big[a^{2}+Q^{2}+r_{h}\left(r_{h}-r_{h}^{-3\omega}\alpha \right) \Big]+\frac{1}{\beta}\Big[2 r_{h}^{1-3\omega}\left(1-\sqrt{1- \frac{\beta}{4r^{2}_{h}}}\right)  \cr&& \times \lbrace r_{h}^{3\omega}\left( a^{2}+Q^{2}-r_{h}^{2}\right)-3r_{h}\alpha\omega \rbrace \Big]
\end{eqnarray}
and 
\begin{eqnarray}\label{eq49}
G_{EUP}&=&\frac{1}{2r_{h}}\Big[a^{2}+Q^{2}+r_{h}\left(r_{h}-r_{h}^{-3\omega}\alpha \right) \Big]+2r^{2}_{h}\Big[ r_{h}-\frac{1}{2r_{h}}\cr&& \times\left( a^{2}+Q^{2}+r_{h}\left(r_{h}-r_{h}^{-3\omega}\alpha \right)\right) -\frac{r_{h}^{-3\omega}\alpha (1-3\omega)}{2}\Big]\left( \frac{1}{4r^{2}_{h}}+\frac{\eta}{L^{2}_{\ast}}\right).
\end{eqnarray}
From the $G$ vs $r_{h}$ graphs shown in Fig. \ref{fig7} and Fig. \ref{fig8} for $\omega = -1/3 $ and $\omega = -2/3 $, we can examine the thermodynamic stability of the KNBHQ. The regions where $ G $ is minimum indicate a more stable state of the black hole, while the regions where $ G $ is relatively high indicate the unstable or less favorable state. The sharp increase or divergence of $ G $  at small $ r_{h} $ in the EUP graph indicates strong instability of small black holes and suppression of tiny horizon radii. Strong modifications are observed in the case of EGUP and EUP corrections, while the GUP correction dominates at small $ r_{h} $, i.e., at the quantum scale. In the case of the EGUP and EUP graphs shown in Fig. \ref{fig7} for $ \omega=-1/3 $, $ G $ drops sharply at large $ r_{h} $, suggesting thermodynamically favored large black holes.

To further study the global stability of the black hole, we investigate the Gibbs free energy as a function of temperature, as illustrated in Figs. \ref{Gibbs13} and \ref{Gibbs23}. This analysis provides a deeper insight into the phase structures suggested by the heat capacity results. For the EGUP and EUP cases with $\omega = -1/3$, where the heat capacity exhibits two divergence points and a region of instability ($C_j < 0$), the Gibbs free energy displays a characteristic swallowtail structure. The intersection point within this swallowtail indicates a first-order phase transition between the small and large black hole phases, while the cusps correspond to the two second-order phase transition points. In contrast, the $G-T$ graphs for the GUP case ($\omega = -1/3$) and all cases for $\omega = -2/3$ show the absence of a swallowtail and the presence of only a single cusp, indicating a single second-order phase transition. Notably, these results are entirely consistent with the heat capacity analysis, where the number of divergence points and the regions of stability ($C_j > 0$) map directly onto the branches and cusps of the Gibbs free energy. Furthermore, the positions of these transitions shift toward higher temperatures as the correction parameters increase, illustrating the impact of quantum gravity effects on the black hole's stability.
\begin{figure}[b!]
    \centering
    \begin{subfigure}{0.45\textwidth}
        \includegraphics[width=\linewidth]{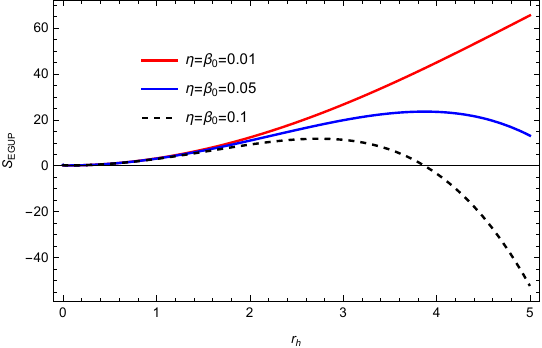}
        \caption{}
    \end{subfigure}
    \hfill
    \begin{subfigure}{0.45\textwidth}
        \includegraphics[width=\linewidth]{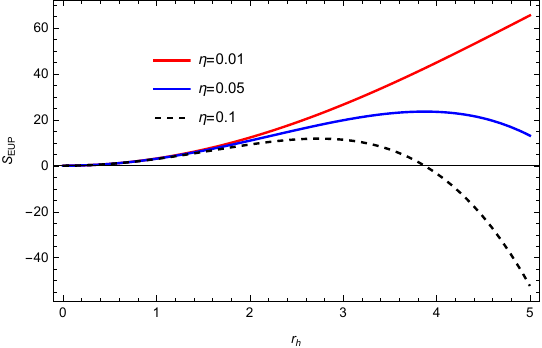}
        \caption{}
    \end{subfigure}
    
    \begin{subfigure}{0.45\textwidth}
        \includegraphics[width=\linewidth]{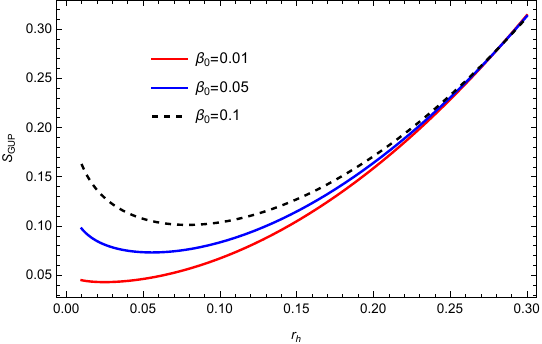}
        \caption{}
    \end{subfigure}
    
    \caption{Entropy as a function of $ r_{h} $ for $ a=0.1 $ and $ L=3 $. (a) EGUP corrected entropy for different $ \eta $ and $ \beta_{0} $. (b) EUP corrected entropy for different $ \eta $. (c) GUP corrected entropy for different $ \beta_{0} $.}
\label{fig.entropy}
\end{figure}
\begin{figure}[b!]
    \centering
    \begin{subfigure}{0.45\textwidth}
        \includegraphics[width=\linewidth]{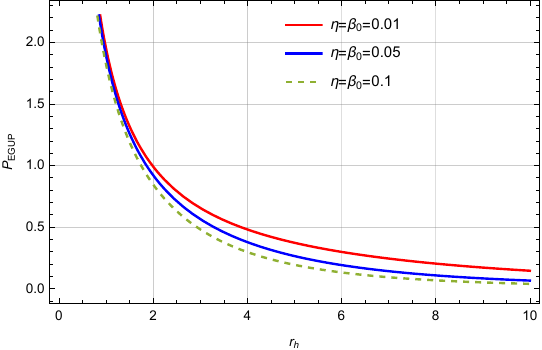}
        \caption{}
    \end{subfigure}
    \hfill
    \begin{subfigure}{0.45\textwidth}
        \includegraphics[width=\linewidth]{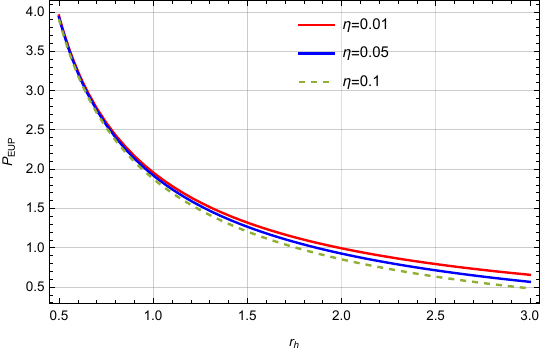}
        \caption{}
    \end{subfigure}
    
    \begin{subfigure}{0.45\textwidth}
        \includegraphics[width=\linewidth]{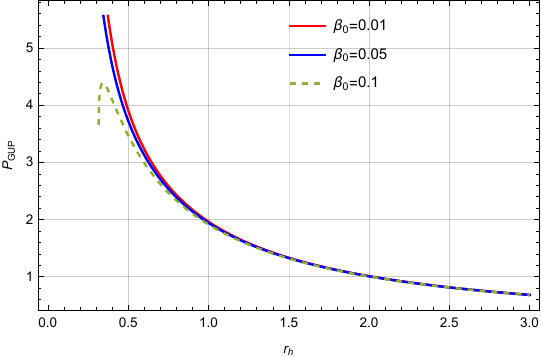}
        \caption{}
    \end{subfigure}
    
    \caption{Pressure as a function of $ r_{h} $ for $ a=0.1 $, $ Q=0.3 $, $ T=1 $, $ L=3 $, and $ \omega=-\frac{1}{3} $. (a) EGUP corrected pressure for different $ \eta $ and $ \beta_{0} $. (b) EUP corrected pressure for different $ \eta $. (c) GUP corrected pressure for different $ \beta_{0} $.}
\label{fig5}
\end{figure}
\begin{figure}[b!]
    \centering
    \begin{subfigure}{0.45\textwidth}
        \includegraphics[width=\linewidth]{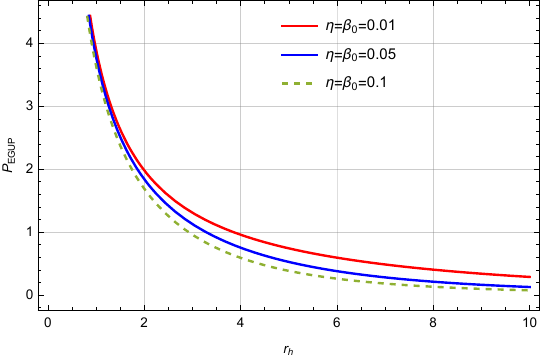}
        \caption{}
    \end{subfigure}
    \hfill
    \begin{subfigure}{0.45\textwidth}
        \includegraphics[width=\linewidth]{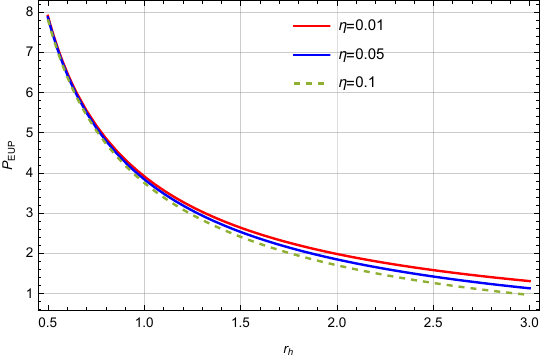}
        \caption{}
    \end{subfigure}
    
    \begin{subfigure}{0.45\textwidth}
        \includegraphics[width=\linewidth]{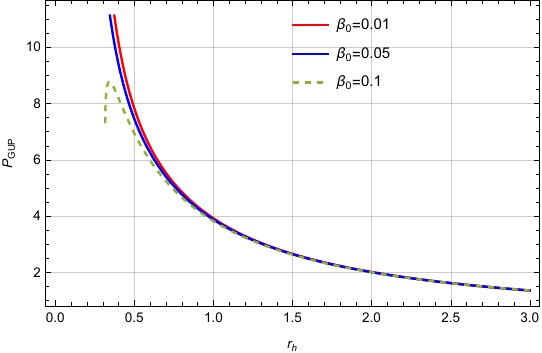}
        \caption{}
    \end{subfigure}
    
    \caption{Pressure as a function of $ r_{h} $ for $ a=0.1 $, $ Q=0.3 $, $ T=1 $, $ L=3 $, and $ \omega=-\frac{2}{3} $. (a) EGUP corrected pressure for different $ \eta $ and $ \beta_{0} $. (b) EUP corrected pressure for different $ \eta $. (c) GUP corrected pressure for different $ \beta_{0} $.}
\label{fig6}
\end{figure}
\subsection{Entropy}

From Eq. (\ref{eq19}), the entropy of the black hole, as modified by EGUP, can be expressed as follows
\begin{eqnarray}\label{eq38}
S_{EGUP}= \int \frac{1}{\frac{32r^{2}_{h}}{\beta}\left\lbrace  1-\sqrt{1-\beta\left(\frac{1}{4r^{2}_{h}}+\frac{\eta}{L^{2}_{\ast}} \right) } \right\rbrace }dA.
\end{eqnarray}
By utilizing the expression of the area of the event horizon of the KNBHQ, given by $ A=4\pi(r^{2}_{h}+a^{2}) $ and applying the Taylor series expansion to Eq. (\ref{eq38}) and ignoring higher-order terms, the integration reduces to:
\begin{eqnarray}\label{eq39}
S_{EGUP}= S-\frac{\eta}{8\pi L_{\ast}}\left(A-4\pi a^{2} \right)^{2}-\frac{\beta \pi}{16}\ln(A-4\pi a^{2}), 
\end{eqnarray}
where $ S=\pi(r^{2}_{h}+a^{2}) $ is the original entropy of the KNBHQ.

Similarly, the entropies modified by GUP and EUP  are given by
\begin{eqnarray}\label{eq40}
S_{GUP}=S-\frac{\beta \pi}{16}\ln(A-4\pi a^{2})
\end{eqnarray}
and
\begin{eqnarray}\label{eq41}
S_{EUP}=S-\frac{\eta}{8\pi L_{\ast}}\left(A-4\pi a^{2} \right)^{2}.
\end{eqnarray}
The changes in entropy with respect to $ r_{h} $ of the KNBHQ, affected by EGUP, EUP and GUP, are illustrated in Fig. \ref{fig.entropy}. The corrected entropies of EGUP, EUP, and GUP appear to be independent of the quintessence parameters $ (\omega,\alpha) $. In EGUP and EUP, entropy rises gradually when the correction parameters are small. However, as the correction strength increases, entropy increases to a peak and then sharply decreases, even becoming negative at large $ r_{h} $.  This indicates a loss of thermodynamic stability. Conversely, for GUP-corrected entropy, it remains positive and increases as $ r_{h} $ increases, indicating that thermodynamic stability is maintained. 

\subsection{Pressure}
We introduce the pressure $ P $ and the matter-energy density $ \rho $ associated with the KNBHQ as\cite{52}
\begin{eqnarray}\label{eq42}
P=-\frac{3\alpha}{2}\frac{\omega^{2}}{r^{3(1+\omega)}}.
\end{eqnarray}
By resolving for $ \alpha $ from Eq. (\ref{eq20}) and inserting it into Eq. (\ref{eq42}), we obtain the pressure adjusted by EGUP as
\begin{eqnarray}\label{eq43}
P_{EGUP}=\frac{\omega}{2r^{5}_{h}}\Big[r_{h}(r^{2}_{h}-a^{2}-Q^{2}) -\frac{\pi \beta T (r^{2}_{h}+a^{2})}{2\left\lbrace  1-\sqrt{1-\beta\left( \frac{1}{4r^{2}_{h}}+\frac{\eta}{L^{2}_{\ast}}\right) } \right\rbrace }\Big].
\end{eqnarray}

When $ \eta=0 $, we get the pressure modified by GUP as
\begin{eqnarray}\label{eq44}
P_{GUP}=\frac{\omega}{2r^{5}_{h}}\Big[r_{h}(r^{2}_{h}-a^{2}-Q^{2}) -\frac{\pi \beta T (r^{2}_{h}+a^{2})}{2\left\lbrace  1-\sqrt{1- \frac{\beta}{4r^{2}_{h}} } \right\rbrace }\Big].
\end{eqnarray}

When $ \beta=0 $, we get the pressure modified by EUP as
\begin{eqnarray}\label{eq45}
P_{EUP}=\frac{\omega}{2r^{2}_{h}}\Big[\frac{(r^{2}_{h}-a^{2}-Q^{2})}{r_{h}}-\frac{4\pi L^{2}_{\ast} T (r^{2}_{h}+a^{2})}{L^{2}_{\ast}+4r^{2}_{h}\eta} \Big].
\end{eqnarray}
To examine how the pressure varies with the position of the event horizon radius, we illustrate the pressures corrected by EGUP, GUP and EUP as a function of the event horizon radius in Fig. \ref{fig5} and Fig. \ref{fig6} for $ \omega=-\frac{1}{3} $ and $ \omega=-\frac{2}{3} $. In this analysis, the parameters utilized include $ a=0.1 $, $ Q=0.3 $, $ T=1 $ and $ L=3 $. The corrected pressures are found to be affected by the quantum correction parameters $ (\eta, \beta) $ and the quintessence field parameter $ \omega $. In all cases, the pressure decreases with increasing $r_h$, showing higher pressure for smaller black holes. Further increasing $\eta$ and $\beta$  lower the pressure. The variation in pressure becomes more significant for larger black holes in the EGUP and EUP cases, whereas the opposite behavior is observed for the GUP case, where the effect is stronger for smaller black holes.

\section{Discussion and Conclusion}
In this study, we explore the impact of the EGUP on the thermodynamic properties of Kerr-Newman black holes when quintessence is taken into account. To begin, we examined different modified uncertainty principles and presented the concept of EGUP. Following this, we analyze how EGUP, along with the GUP and the EUP, modifies the Hawking temperature in the presence of quintessence. Based on Figs. \ref{fig1} and \ref{fig2}, it can be observed that $ T_{EGUP} $, $ T_{EUP} $ and $ T_{GUP} $ yield positive real values when  $r_{h}\geq0.330$ with $\omega=-\frac{1}{3} $. However, when $ \omega=-\frac{2}{3} $, $ T_{EGUP} $, $ T_{EUP} $, and $ T_{GUP} $ show positive and real values within the range of $0.320 < r_{h} < 5$. EGUP and EUP influence the Hawking temperature to some extent, whereas GUP's effect on the Hawking temperature is negligible. In the analysis involving EGUP and EUP corrected heat capacity displayed in Figs. \ref{fig3b} and \ref{fig3c} reveals two distinct divergence points for $ \omega=-\frac{1}{3} $, whereas for $ \omega=-\frac{2}{3} $, only a single phase is evident, indicating the presence of second-order phase transition. Base on the divergence points, we can divide the system into three thermodynamic regimes: a small black hole phase and a large black hole phase, both characterized by positive heat capacity and hence thermodynamic stability, and an intermediate phase with negative heat capacity, indicating instability. However, in the case of GUP corrected heat capacity, a single phase transition occurs in both scenarios.  EGUP and EUP corrected heat capacity of black holes is negative only in a small range and is positive almost everywhere, showing that there is a more stable region than GUP. Next, we examine the EGUP and GUP remnant temperature and mass of the KNBHQ. In the EUP scenario, we observe that there is no remnant temperature and mass that is dependent on $ \eta $. Then, we study the Gibbs free energy of KNBHQ which plays a central role in determining the thermodynamic stability and phase structure of the black hole. The sharp increase or divergence of $ G $  at small $ r_{h} $ in the EUP graph indicates strong instability of small black holes and suppression of tiny horizon radii. Strong modifications are observed in the case of EGUP and EUP corrections, while the GUP correction dominates at small $ r_{h} $. In the case of the EGUP and EUP graphs shown in Fig. \ref{fig7} for $ \omega=-1/3 $, $ G $ drops sharply at large $ r_{h} $, suggesting thermodynamically favored large black holes. The behavior of the Gibbs free energy with temperature further reveals the nature of phase transitions. For $ \omega=-1/3 $, a characteristic swallowtail structure appears, signaling a first-order phase transition, with cusp points marking second-order transitions. In contrast, for $ \omega=-2/3 $, the absence of a swallowtail and the presence of only a single cusp indicate a purely second-order phase transition. These findings are consistent with heat capacity analysis, and overall, the results demonstrate that quantum gravity corrections significantly modify stability and shift phase transitions to higher temperatures. Furthermore, our calculations reveal that the corrected entropies for EGUP, EUP, and GUP are not influenced by the quintessence field parameter $ \omega $. When the correction parameters are small, entropy rises smoothly. But when the correction parameters are large, entropy rises to a peak and decreases gradually which indicates thermodynamics instability. In the case of GUP, entropy remains positive and increases as $ r_{h} $ increases, indicating thermodynamic stability. Finally, we present a detailed analysis of the EGUP, EUP and GUP corrected pressure as a function of $ r_{h} $ under isothermal conditions. From Figs. \ref{fig5} and \ref{fig6}, we found that the corrected pressures increase as the quintessence field parameter $ \omega $ decreases.

\section*{Acknowledgments}
The first author is being supported by the INSPIRE Fellowships of the Department
of Science and Technology (DST), New Delhi, India (INSPIRE Code: IF200576).

\section*{ORCID}

\noindent Aheibam Boycha Meitei - \url{https://orcid.org/0009-0001-7566-5061}

\noindent Yenshembam Priyobarta Singh - \url{https://orcid.org/0000-0003-3168-7493}

\noindent Telem Ibungochouba Singh - \url{https://orcid.org/0000-0002-2568-0343}

\noindent Irom Ablu Meitei - \url{https://orcid.org/0000-0001-7420-7774}

\noindent Kangujam Yugindro Singh - \url{https://orcid.org/0000-0002-8976-8133}

\end{document}